 \documentclass[useAMS,usenatbib]{mn2e}

\usepackage{graphicx}
\title[ Low angular momentum  flow model I\hspace{-.1em}I for Sgr A* ]
 { Low angular momentum flow model I\hspace{-.1em}I for Sgr A* }
\author[T. Okuda] {T. Okuda
 \thanks{E-mail:bbnbh669@ybb.ne.jp} \\ Nishi-Asahioka-Cho 3-15-1, Hakodate 042-0915, Hokkaido, Japan}

\begin{document}

\date{Accepted }

\pagerange{\pageref{0}--\pageref{0}} \pubyear{2012}

\maketitle

\label{firstpage}

\begin{abstract}
 
 We examine 1D two-temperature accretion flows around a supermassive black hole,
 adopting the specific angular momentum $\lambda$, the total specific energy $\epsilon$ 
 and the input accretion rate $\dot M_{\rm input} = 4.0\times 10^{-6}\;{\rm M_{\odot}}$ yr$^{-1}$ 
 estimated in the recent analysis of stellar wind of nearby stars around Sgr A*.
 The two-temperature flow is almost adiabatic even if we take  account of the heating of
  electrons by ions, the bremsstrahlung cooling and the synchrotron cooling, as long as
 the ratio $\beta$ of the magnetic energy density to the thermal energy density is taken
 to be as $\beta \le 1$.  The different temperatures of ions and electrons are caused by
 the different adiabatic indices of ions and electrons which depend on their temperature
 states under the relativistic regime. The total luminosity increases with increasing 
 $\beta$ and results in $\sim 10^{35}$ -- $10^{36}$ erg s$^{-1}$ for $\beta=10^{-3}$ - 1.
 Furthermore, from 2D time-dependent hydrodynamical calculations of the above flow,
 we find that the irregularly oscillating shocks are formed in the inner region
 and that the luminosity and the mass-outflow rate vary
 by a factor of 2 -- 3 and 1.5 -- 4, respectively. The time variability may be relevant to
 the flare activity of Sgr A*.

\end{abstract}


\section{Introduction}

Sgr A* in our Galactic Centre has been extensively studied in the 
category of accretion processes because it is a supermassive black hole
 candidate in our Galaxy and has unique observational features incompatible
 with the standard thin disc model (Shakura-Sunyaev, hereafter S-S,
 model; Shakura \& Sunyaev 1973).
One of the remarkable features of Sgr A* is that the observed luminosity 
is five orders of magnitude lower than that predicted by the S-S model.
Moreover, the spectrum of Sgr A* differs from the multi-temperature blackbody
spectrum obtained from the S--S model.
Since the observational features of Sgr A* cannot be explained by the S--S
model, two types of theoretical models, namely the  spherical Bondi accretion
model without any net angular momentum \citep{b2} and the advection-dominated 
accretion flow (ADAF) model with high angular momentum \citep{b17,b18}, 
have been proposed (see Narayan \& McClintock 2008, Yuan 2011 and Yuan \& Narayan 2014
  for review). 
Both the Bondi model and the ADAF model result in highly advected flows and
 the radiative efficiency is so low as to be compatible with the observations.
However, in contrast with the simple Bondi model,  the ADAF models were 
shown to be generally successful and more advanced models
\citep{b26,b27}, taking into account the parametric description of the outflow
 and jet, explain well the observations.
The important key to these models for Sgr A* is the amount of angular 
momentum in the accretion flow.
However, at present, we have no clear evidence for the angular momentum from
observations.

The low angular momentum flow model belongs to an intermediate case between the 
Bondi model and the ADAF model and was applied to Sgr A* \citep {b13,b4}. 
Assuming that the Wolf-Rayet star ${\rm IRS}$ 13 $ \rm E3$ is the dominant
source of the matter accreting on to Sgr A* and assuming the wind
temperature $T_{\rm wind}$ = 1.0 or 0.5 keV, they estimated the net 
angular momentum  $\lambda$ of 1.68--2.16 and the Bernoulli constant 
$\epsilon$ of $3.97 \times 10^{-6}$--$1.98 \times 10^{-6}$, 
where the mass $M$ of Sgr A*, the speed of light $c$ and the Schwarzschild radius 
$R_{\rm g}=2GM/c^2$ are used as the units of mass, velocity and distance
and $G$ is the gravitational constant. 
With these flow parameters for $\lambda$ and $\epsilon$,
they showed analytically that there is no continuous flow solution which 
attains to the event horizon, and the resulting flow would be non-stationary, 
but that, for the case of the angular momentum $\lambda$= 1.55 lower than 
the best estimates for Sgr A*, there exists a standard stationary shock
 solution. 
Motivated by their suggestion and results, we examined the low angular 
momentum flow model for Sgr A* using 2D time-dependent hydrodynamical 
calculations and discussed the results on the activity of Sgr A* \citep{b19}.
However, in the initial model used, we assumed a constant ratio of the ion 
temperature $T_{\rm i}$ to the electron temperature $T_{\rm e}$ throughout
the region.
In this paper, without using such assumption, we solve coherently relevant
 differential equations of the accretion flow which include the radiative processes 
of heating and cooling, and examine time-dependent behaviours of the
2D accretion flow in the relevance to the activity of Sgr A*. 

\section{Stationary low angular momentum accretion flow}
\subsection{Modeling a 1D low angular momentum flow}

 We consider a supermassive black hole with mass $M= 4\times 10^6 M_{\odot}$
 and examine a 1D low angular momentum flow around  Sgr A*.
 We use here  typical flow parameters of the specific angular momentum $\lambda$, 
 the specific total energy $\epsilon$ and the mass accretion rate $\dot M 
  =4.0 \times 10^{-6}M_{\odot}$ yr$^{-1} $which were estimated for the accretion flow
 around  Sgr A*  \citep{b13}.

 First, we solve the Bernoulli equation of a thin, rotating,
 inviscid and adiabatic accretion flow with a single temperature $T$ and a constant
 angular momentum $\lambda$, find the outer and inner sonic
 points, and get the Mach number versus radius relation, the sound speed 
$v_{\rm s}$, the thickness $h$ of the accretion flow, the radial velocity $v$
 and the temperature $T$ at a given radius $r$ as is done in \citet{b3}.
 We assume here that  the accretion flow is in hydrostatic equilibrium in 
 vertical direction
\begin{equation}
  {p\over h} = \rho { GMh \over {(r-R_{\rm g})^2 r}},
\end{equation}
where $p$ and $\rho$ are the total gas pressure and the density.
The Bernoulli constant $\epsilon$ is given by
 
 \begin{equation}
  \epsilon= {1\over 2} v^2+ {1\over {\gamma-1}} {v_{\rm s}}^2 -
  {GM\over {r-R_{\rm g}}} + {1\over 2} {\lambda^2 \over r^2},
\end{equation}
where  $v_{\rm s}$ and  $\gamma$  are  the sound velocity and 
the specific heat ratio. 
In this paper, we consider two models of $\lambda$ = 1.68 and $\epsilon$ = $3.97 
\times 10^{-6}$ (model A) and $\lambda$ = 1.35 and $\epsilon$ = $1.98 
\times 10^{-6}$ (model B) in the usual nondimensional units.
The Mach number versus radius relation in the models 
 is given in figs~1 and~2 in the previous paper \citep{b19}. 
In model A, the particle which passes through the outer sonic point falls 
down supersonically inwards but never attains the event horizon since it
makes a closed loop of the Mach number curve.
On the other hand, in model B, the particle falls supersonically, 
jumps to a subsonic state at the shock position of $R_{\rm s} \sim 20 R_{\rm g}$ and 
tends supersonically towards the event horizon.

\begin{table*}
\centering
\caption{Model parameters of the specific angular momentum $\lambda$, the 
 specific total energy $\epsilon$ and the adiabatic index  $\gamma_{\rm i}$  
of ions for the accreting matter on to Sgr A*, where
 the radial velocity $v_{\rm out}$, the Mach number $M_{\rm a}$, 
the ion temperature $T_{\rm i}$, the difference $T_{\rm i}-T_{\rm e}$
 between the ion temperature  $T_{\rm i}$ and the electron temperature $T_{\rm e}$,
 the ratio $\beta$ of the magnetic energy density to the thermal energy density 
 at $R_{\rm out}=10^3 R_{\rm g}$ in the two-temperature model
 are also shown.}
\begin{tabular}{@{}cccccccccc} \hline \hline
Model&$\lambda$ & $\epsilon$& $\gamma_{\rm i}$& $\dot M$ $(M_{\odot}$
${\rm yr}^{-1})$ &$v_{\rm out}/c$ & $M_{\rm a}$ & $T_{\rm i}$ (K)
  & $(T_{\rm i}-T_{\rm e})$ (K) & $\beta$               \\      \hline
 $A$    & 1.68  & 3.97 $\times 10^{-6}$ & 1.6& 4.0 $\times 10^{-6}$
        & $1.971\times 10^{-2}$ &1.3729  & $2.824\times 10^8$ &$870$& $10^{-3}$ -- 1 \\
 $B$    & 1.35  & 1.98 $\times 10^{-6}$ & 1.6  &4.0 $\times 10^{-6}$
        & $1.989\times10^{-2}$ & 1.5197 & $ 2.606\times 10^8$  & 770 & $10^{-3}$ -- 1 \\
\hline
\end{tabular}
\end{table*}

Secondly, we adopt a two-temperature model of the accretion flow
with the ion temperature $T_{\rm i}$ and the electron temperature $T_{\rm e}$, taking
account of the radiative processes of heating of electrons by ions, 
bremsstrahlung  cooling and synchrotron cooling in the optically thin limit.
From the 1D  stationary equations of the inviscid and non-adiabatic 
accretion flow with the constant angular momentum, we have the following differential equations of $T_{\rm e}$, 
$T_{\rm i}$ and $\eta =v^2/({p\over\rho})$ \citep{b14}:

\begin{equation}
 {{d{\rm ln}T_{\rm e}} \over  dr}=(\gamma_{\rm e}-1){{A(\eta-1)+(\gamma_{\rm i}-1)\Gamma_{\rm i}({1\over 2}\eta-1)(A-B)-C}
 \over {\eta-1+({1\over 2}\eta-1)(\gamma_{\rm e}\Gamma_{\rm e}+\gamma_{\rm i}\Gamma_{\rm i}-1)}}
 ,
 \end{equation}

\begin{equation}
 {{d{\rm ln}T_{\rm i}} \over dr}=(\gamma_{\rm i}-1){{B(\eta-1)-(\gamma_{\rm e}-1)\Gamma_{\rm e}({1\over 2}\eta-1)
 (A-B)-C} \over {\eta-1+({1\over 2}\eta-1)(\gamma_{\rm e}\Gamma_{\rm e}+\gamma_{\rm i}\Gamma_{\rm i}-1)}},
\end{equation}

\begin{equation}
 {{d{\rm ln}\eta} \over dr}=2\left[A-({1\over{\gamma_{\rm e}-1}}+\Gamma_{\rm e}){{d{\rm ln}T_{\rm e}} \over  dr}
 -\Gamma_{\rm i}{{d{\rm ln}T_{\rm i}} \over dr}\right],
\end{equation}

\begin{eqnarray}
 A &=& {{4{\rm \pi}\left({ {GM\over r} R_{\rm G}T_{\rm e}\Gamma_{\rm e}}\right)^{-1/2}(r-R_{\rm g})r}\over
 {\dot M}}(q^{\rm ie}-q_{\rm br}-q_{\rm syn}) \nonumber \\
 &&\;\; - {3\over 2r}- {1\over {r-R_{\rm g}}},
\end{eqnarray}

\begin{equation}
B={{4{\rm \pi} \left({ {{GM\over r}R_{\rm G}T_{\rm i}\Gamma_{\rm i}}}\right)^{-1/ 2}(r-R_{\rm g})r}\over
 {\dot M}} q^{\rm ie} -{3\over 2r}- {1\over {r-R_{\rm g}}},
\end{equation}
\begin{equation}
C= -{{({\Omega_{\rm k}}^2-\Omega^2) r} \over {R_{\rm G}(T_{\rm e}+T_{\rm i})}}
 + {3\over 2r} + {1 \over {r-R_{\rm g}}}, 
\end{equation}
where $p = p_{\rm i}+p_{\rm e}$, $p_{\rm i}$ and $p_{\rm e}$ are the ion and the electron gas pressure,
  $R_{\rm G}$ is the gas constant,  $\Omega$ is the angular velocity, $\Omega_{\rm K}$ is the
 Keplerian angular velocity, $\dot M$(=4${\rm \pi} rh\rho v$) is the mass accretion rate,
  $\Gamma_{\rm e}= T_{\rm e}/ (T_{\rm e}+T_{\rm i})$ and
 $\Gamma_{\rm i}= T_{\rm i} /(T_{\rm e}+T_{\rm i})$.
 $\gamma_{\rm i}$ and $\gamma_{\rm e}$ are the adiabatic indices of ions and electrons, respectively, 
 which may differ a little depending on their temperature states and the different indices cause
 different temperatures of electrons and ions even in the adiabatic state as is found from equations (3) and (4) \citep{b25}.
 Here, we use that $\gamma_{\rm e}$ is 1.6 at $kT_{\rm e} \le m_{\rm e} c^2$, 
 that is, $T_{\rm e} \le 5.9\times 10^9 {\rm K}(=T_{\rm c})$ where electrons are
 non-relativistic but is 4/3 at $kT_{\rm e} \ge m_{\rm e} c^2$ where electrons become relativistic,
 while $\gamma_{\rm i}$ is taken to be 1.6 throughout the region because ions remain in a 
 non-relativistic state of $kT_{\rm i} \le m_{\rm p}c^2$ \citep{b7}. $T_{\rm c}$ is the critical electron
 temperature at which electrons in the non-relativistic state change into the relativistic state.
 $q^{\rm ei}$ and $q_{\rm br}$ are the energy transfer rate from ions to  electrons by Coulomb collisions
 and the cooling rate of electrons by electron-ion and electron-electron bremsstrahlungs 
 and are given as follows \citep{b22}.
   \begin{eqnarray}
 &q^{\rm ie}&=5.61\times 10^{-32} {{n_{\rm e}n_{\rm i}(T_{\rm i}-T_{\rm e})}\over {K_2(1/\theta_{\rm e})
  K_2(1/\theta_{\rm i})}} \nonumber \\
  && \times \left[{{2(\theta_{\rm e}+\theta_{\rm i})^2+1}\over (\theta_{\rm e}+\theta_{\rm i})}
  K_1({{\theta_{\rm e}+\theta_{\rm i}}\over {\theta_{\rm e}\theta_{\rm i}}}) 
  +2K_0({{\theta_{\rm e}+\theta_{\rm i}}\over {\theta_{\rm e}\theta_{\rm i}}})\right] \nonumber \\
   &&  \hspace{3.8cm} {\rm erg \ cm^{-3} \; s^{-1}},
 \end{eqnarray}
   \begin{equation}
  q_{\rm br} = q_{\rm ei} + q_{\rm ee},
  \end{equation}
  \begin{equation}
  q_{\rm ei}=1.48\times 10^{-22}n_{\rm e}^2 F_{\rm ei}(\theta_{\rm e}) \hspace{1cm}{\rm erg \ cm^{-3}\; s^{-1}},
  \end{equation}
  \begin{eqnarray}
   \;\;\;F_{\rm ei}(\theta_{\rm e}) =\left\{\begin{array}{ll}
    1.02\theta_{\rm e}^{1/2}(1+1.78\theta_{\rm e}^{1.34}) & \mbox{ for $\theta_{\rm e} <1$},  \\
    1.43\theta_{\rm e}[{\rm ln}(1.12\theta_{\rm e}+0.48)+1.5]
                                                          & \mbox{ for  $\theta_{\rm e}>1$},
                                              \end{array}
                                      \right. 
  \end{eqnarray}
  
  \begin{eqnarray}
  q_{\rm ee} = \left\{\begin{array}{llll}
     2.56\times 10^{-22}n_{\rm e}^2\theta_{\rm e}^{1.5}
       (1+1.10\theta_{\rm e}+\theta_{\rm e}^2-1.25\theta_{\rm e}^{2.5})  & \\
     \hspace{3cm}  {\rm erg \ cm^{-3} \; s^{-1} }  \mbox{ for $\theta_{\rm e} <1$}, 
  & \\
    3.40\times 10^{-22}n_{\rm e}^2\theta_{\rm e} [{\rm ln}(1.123\theta_{\rm e})+1.28 ]
  & \\                                                                                       
     \hspace{3cm}   {\rm erg \ cm^{-3} \; s^{-1}}  \mbox{ for  $\theta_{\rm e} >1$}, & \\
                               \end{array}
                      \right.
    \end{eqnarray}
  where $n_{\rm e}$ and $n_{\rm i}$ are the number density of electrons and ions, $K_0$, $K_1$
  and $K_2$ are
 modified Bessel functions, and the dimensionless electron and ion temperature are defined by

  \begin{equation}
  \theta_{\rm e}= {kT_{\rm e}\over {m_{\rm e}c^2}}, \;\;\; \theta_{\rm i}={kT_{\rm i}\over {m_{\rm p}c^2}}.
  \end{equation}
  $q_{\rm syn}$ is the  synchrotron cooling rate which is given by \citep{b17,b6}

  \begin{eqnarray}
  q_{\rm syn}  &=& 
    {{2{\rm \pi} kT_{\rm e}\nu_{\rm c}^3 }\over {3hc^2}}  
     +6.76\times 10^{-28} {n_{\rm i}\over{K_2(1/\theta_{\rm e})a_1^{1/6}}} \nonumber \\ 
  & \times &  [{1\over a_4^{11/2}}\Gamma({11\over 2},a_4\nu_{\rm c}^{1/3})
     +{a_2\over a_4^{19/4}} \Gamma({19\over 4},a_4\nu_{\rm c}^{1/3}) \nonumber \\        
  &+& {a_3\over a_4^4}( a_4^3\nu_{\rm c}+3a_4^2\nu_{\rm c}^{2/3}    
   +6a_4\nu_{\rm c}^{1/3}+6) {\rm e}^{-a_4\nu_{\rm c}^{1/3}}] \nonumber \\   
  && \hspace{3.8cm} {\rm erg \ cm^{-3} \ s^{-1}},
   \end{eqnarray}
   where 
    \begin{eqnarray}
   a_1={2\over {3\nu_0\theta_{\rm e}^2}},\; a_2={0.4\over a_1^{1/4}},\; a_3={0.5316\over a_1^{1/2}},
   \; a_4={1.8899a_1^{1/3}}, \nonumber \\                                    
   \Gamma(a,x)= \int_{x}^{\infty}t^{a-1}{\rm e}^{-t}dt, 
   \nu_0={eB\over {2{\rm \pi} m_{\rm e}c}} {\; \rm and} \;\nu_{\rm c}={3\over 2}\nu_0\theta_{\rm e}^2x_{\rm M}.
     \end{eqnarray}
     Here $B$ is the strength of the magnetic field and $x_{\rm M}$  is determined from the next equation
     as
    \begin{eqnarray}
      {\rm exp}(1.8899x_{\rm M}^{1/3} ) &=&  2.49\times 10^{-10}{4{\rm \pi} n_{\rm e}r\over
       B} {1\over {\theta_{\rm e}^3 K_2(1/\theta_e) }} \nonumber \\
       &\times&    \left({1\over  x_{\rm M}^{7/6} }
        + {0.40 \over x_{\rm M}^{17/12}}+{0.5316\over {x_{\rm M}^{5/3}}}\right). \nonumber \\        
    \end{eqnarray}
 We give the magnetic field $B$ at each radius, assuming that the ratio $\beta$ 
 of the magnetic energy density to the thermal energy density is 
constant throughout the region. 
 The cooling rates  due to  the pair annihilation processes are neglected here because the equilibrium
  pair number density are found to be much smaller than the ion density in the present temperature region.
 
 Finally, we set the outer boundary radius $R_{\rm out}$  to be $10^3R_{\rm g}$ and 
 determine $T_{\rm e}$, $T_{\rm i}$ and $\eta$ using the adiabatic solutions
 so that the heating rate of electrons by ions equals to the bremsstrahlung
 cooling rate at $r=R_{\rm out}$ and the total gas pressure $p=R_{\rm G}\rho(T_{\rm e}+T_{\rm i})$ has the same
 value as the adiabatic case. 
 The flow variables of models A and B at the outer boundary
 are listed in Table 1.

\subsection{Numerical results}

 \begin{figure}
     \includegraphics[width=86mm,height=66mm]{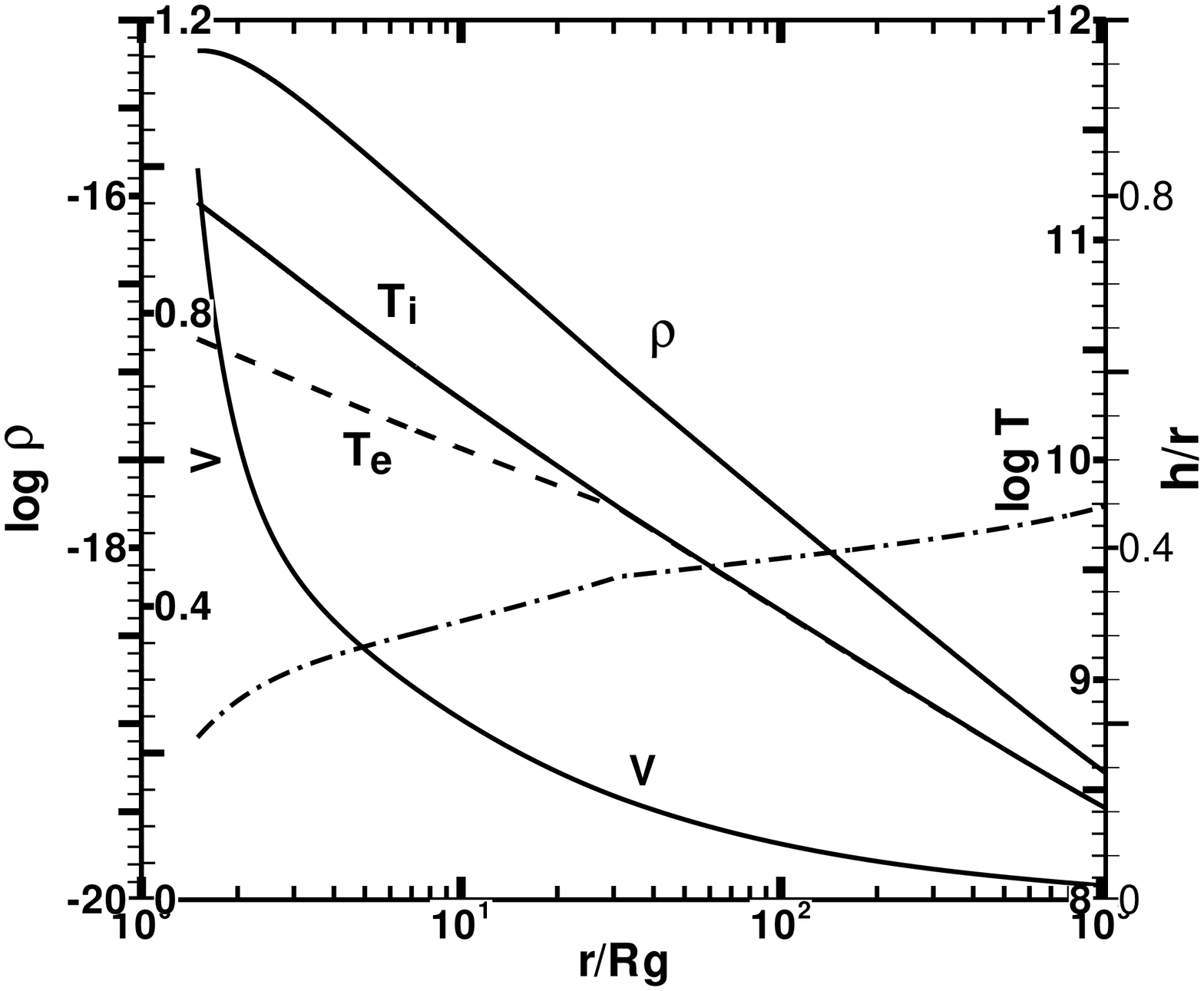}
     \caption{The radial velocity $v$,  the ion temperature $T_{\rm i}$, the electron temperature
 $T_{\rm e}$ (dashed line), the density $\rho$ and the relative half thickness $h/r$ of the accretion flow
 (dash--dotted line) for two--temperature model of model B with $\beta=10^{-3}$.
 The distributions of these variable except $T_{\rm e}$ are almost same as those in one temperature model
  with $T_{\rm i}=T_{\rm e}$ }  
     \label{fig1}
 \end{figure}

 \begin{figure}
     \includegraphics[width=86mm,height=66mm]{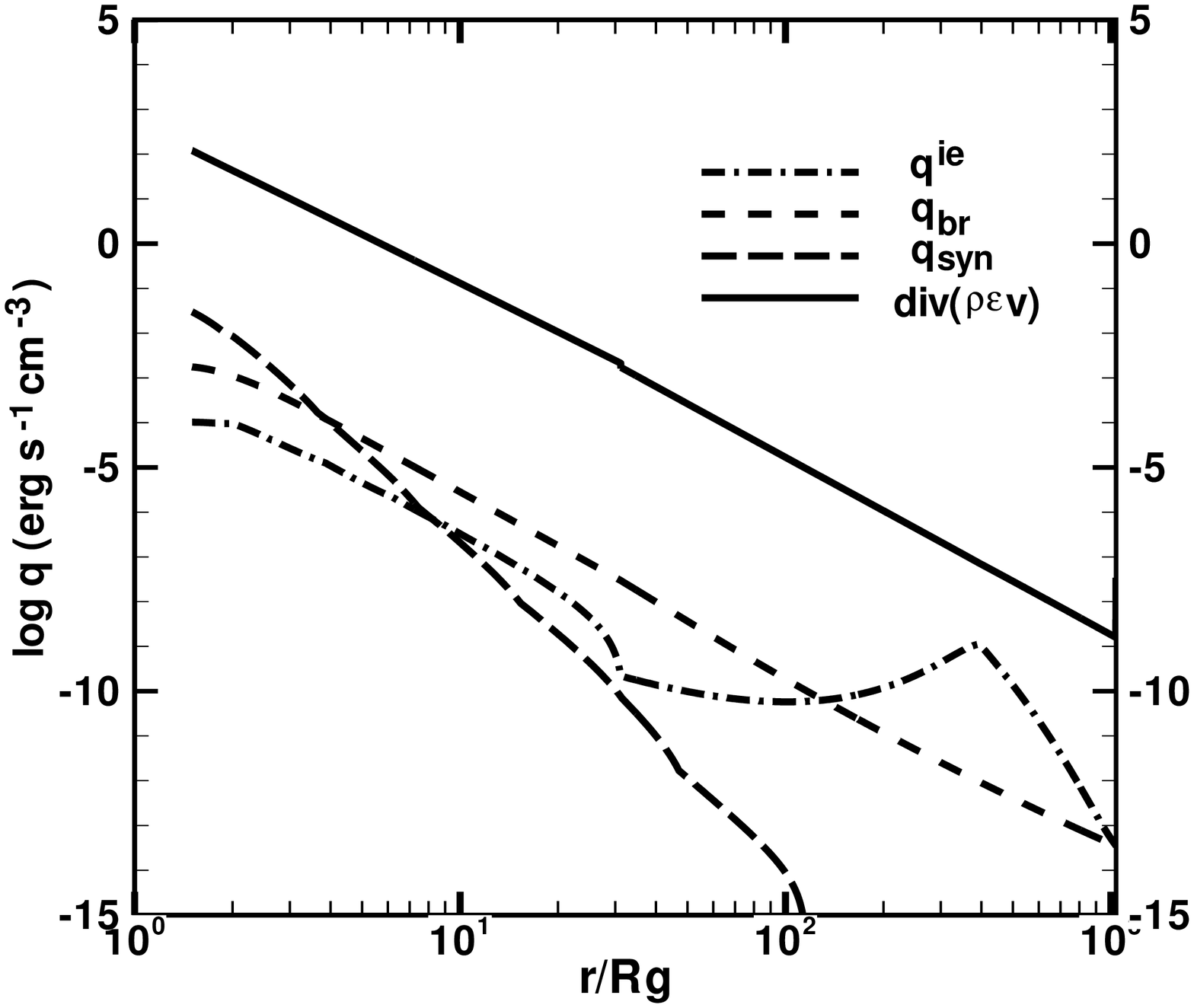}
     \caption{The energy transfer rate $q^{\rm ie}$(dash--dotted line)  from  ions to  electrons by Coulomb collisions,
     the  bremsstrahlung cooling rate $q_{\rm br}$ (dashed line), the synchrotron cooling rate 
     $q_{\rm syn}$ (longdashed lines) and the transfer rate ${\rm div}(\rho \varepsilon {\bmath{v}})$ 
  of the advected thermal energy versus radius in model B with $\beta=10^{-3}$.} 
     \label{fig2}
 \end{figure}

We solve equations (3) -- (5) inwards starting from the outer boundary using a second-order Runge--Kutta method. 
In model A with $\beta= 10^{-3}$ and 1.0, we could not obtain the continuous flow solution which attains the inner boundary  $R_{\rm in}= 1.5 R_{\rm g}$ and
 the solution ended at $r \sim 4.2 R_{\rm g}$.
 If the effects of the radiative energy loss and gain are negligible,
 this is expected from the result of model A in the previous paper \citep{b19},
 because the Mach number versus radius curve in the case would make a closed loop and never attain the event horizon.
While, differently from model A, model B with $\beta=10^{-3}$ -- 1.0 has the continuous solution to the inner edge. Fig.~1 shows the physical variables of the Mach number  $M_{\rm a}$ of the radial velocity $v$, the ion temperature $T_{\rm i}$, the electron temperature $T_{\rm e}$, the density $\rho$ and the half thickness of the accretion
 flow versus radius for model B with $\beta=10^{-3}$.
 The electron temperatures deviate from the ion temperatures at $r/r_{\rm g} 
 \le 30$ and the ratio $T_{\rm i}/T_{\rm e}$ is $\sim 4$ at the inner edge.
 Table 2 shows the total integrated emission $\int q^{\rm ie} {\rm d} V$, $\int q_{\rm br}
 {\rm d}V$ and $\int q_{\rm syn} {\rm d} V$ for model B with $\beta=10^{-3}$ -- $1.0$.
 The synchrotron cooling rate depends on the magnetic field,  that is,
  $B$ and increases with increasing $\beta$, as is found from equations (15) and (16).
 As the result, we obtain total luminosities of $8.7\times 10^{34}$ -- $3.7\times 10^{36}$ erg s$^{-1}$ 
 for model B with $\beta=10^{-3}$ -- $1.0$. 
 
  \begin{table*}
\centering
\caption{The integrated energies of $q^{\rm ie}$, $q_{\rm br}$ and  $q_{\rm syn}$, respectively,
 by the Coulomb collision, the  bremsstrahlung and the synchrotron for model B 
with $\beta=10^{-3}$ -- $1.0 $.}
\begin{tabular}{@{}ccccc} \hline \hline
$\beta$    & Coulomb collision (erg ${\rm s}^{-1}$)& Bremsstrahlung (erg ${\rm s}^{-1}$)& Synchrotron (erg ${\rm s}^{-1}$) \\  \hline
 $1.0$     &  2.1$\times 10^{35}$ &  8.6$\times 10^{34}$  &  3.6$\times 10^{36}$       \\
 $0.1$     &  2.1$\times 10^{35}$ &  8.6$\times 10^{34}$  &  2.2$\times 10^{35}$       \\
 $10^{-2}$ &  2.1$\times 10^{35}$ &  8.6$\times 10^{34}$  &  1.3$\times 10^{34}$       \\
 $10^{-3}$ &  2.1$\times 10^{35}$ &  8.6$\times 10^{34}$  &  7.4$\times 10^{32}$       \\
\hline
\end{tabular}
\end{table*}

 It should be noticed that the different temperatures of ions
 and electrons are caused only by the different adiabatic indices of ions
 and electrons and not by the effects of the radiative energy loss and gain. 
The transfer rate of the advected thermal energy fallen
 into the event horizon is much larger than the total amounts of the radiative heating
 and cooling.  Fig.~2  shows the energy transfer rate $q^{\rm ie}$ (dash--dotted line) from ions to electrons, the  bremsstrahlung cooling rate $q_{\rm br}$ (dashed line), the synchrotron cooling rate  $q_{\rm syn}$ (longdashed line) and the transfer rate $ {\rm div}(\rho \varepsilon {\bmath v})$ of the advected thermal energy versus radius
 in model B.  
 The advected thermal energy term  $ {\rm div}(\rho \varepsilon {\bmath v})$
  is by more than three orders of magnitude larger in the inner region compared 
 with the cooling terms and  balances exactly the rate $-p {\rm div} {\bf v}$ of the internal energy increment by compression.
 Only the synchrotron cooling rate increases with increasing $\beta$ and becomes comparable to the advected thermal energy term  for the case of large $\beta( > 1)$.
 The adiabatic flow is due to the large radial velocity comparable to the free-fall velocity, the too low density
  of the gas and the large accretion flow thickness with $h/r \sim 1$.
 Accordingly, as long as the magnetic field is taken to be as $\beta \le 1$, the effects of the radiative heating and cooling are negligible and the two-temperature model is ascribed to the adiabatic two-temperature model.
 
  Thus, solving the 1D stationary differential equations, we get the accretion flow of model B but not of model A.
  However, we do not know yet whether the shock phenomena found in the previous paper occur or not and 
  what accretion flow occurs actually in model A. The 1D flow equations also do not take account of the outflow
  which plays an important role in the actual accretion flow.
 Then, we examine the time-dependent 2D accretion flow.

\section{ Time-Dependent behaviours of 2D Two-Temperature Model}
\subsection{Basic equations and methods}
 The set of relevant time-dependent equations are given in the spherical
polar coordinates ($r$,$\zeta$,$\varphi$):

\begin{equation}
  { \partial\rho\over\partial t} + {\rm div}(\rho\bmath{v}) =  0,  
\end{equation}
\begin{equation}   
 {\partial(\rho v)\over \partial t} +{\rm div}(\rho v \bmath{v})  =
 \rho\left[{w^2\over r} + {v_\varphi^2\over r}-{GM \over (r-R_{\rm g})^2} 
 \right] -{\partial p\over \partial r},
\end{equation}
\begin{equation}  
 {{\partial(\rho rw)}\over \partial t} +{\rm div}(\rho rw\bmath{v}) 
 = -\rho v_\varphi^2{\rm tan}\zeta-{\partial p\over\partial\zeta}, 
\end{equation}

\begin{equation}    
{{\partial(\rho r{\rm cos}\zeta v_\varphi)}\over \partial t} 
    +{\rm div}(\rho r{\rm cos}
\zeta v_\varphi\bmath{v}) = 0, 
\end{equation}

\begin{equation}  
 {{\partial \rho\varepsilon_{\rm i}}\over \partial t}+
   {\rm div}(\rho\varepsilon_{\rm i}\bmath{v})
     = -p_{\rm i}\;\rm div \bmath{v}  - q^{\rm ie}, 
\end{equation}
and
\begin{equation}  
 {{\partial \rho\varepsilon_{\rm e}}\over \partial t}+
   {\rm div}(\rho\varepsilon_{\rm e}\bmath{v})
     = -p_{\rm e}\;\rm div \bmath{v} + q^{\rm ie} - q_{\rm syn}- q_{\rm br}, 
\end{equation}
where  $\bmath{v} =(v, w, v_\varphi)$ are the
three velocity components,  $\varepsilon_{\rm i}$ and $\varepsilon_{\rm e}$ are the specific
 internal energy of the ion and electron. 
Here we neglect the radiation transport assuming that the flow is optically thin.

The set of partial differential equations (18)--(23) is 
numerically solved by a finite-difference method under adequate initial 
and boundary conditions.
The initial flow variables  are given from the 1D solutions in Section 2
but, at $r \le 4.2R_{\rm g}$ in model A, are given by  extrapolation of nearby variables 
because the 1D solution of model A was obtained only at $4.2 \leq r/R_{\rm g} \leq 10^3$.
 This approximation of the variables never influence the final result
 as long as the flow variables at the outer boundary are fixed constantly and a sufficiently large
 integration time is taken.
The initial atmosphere above the accreting matter in the computational region  is  given 
 as a radially hydrostatic equilibrium state with zero azimuthal velocities everywhere.
Physical variables at the inner boundary, except for the velocities, are given by extrapolation of
 the variables near the boundary. 
 However, we impose limit conditions that the radial velocities at the boundary are given 
 by a free-fall velocity and the angular velocities are zero.
As for the outer boundary region above the outer accreting matter, we  use
free-floating conditions and allow for the outflow of matter, whereas here any 
inflow is prohibited.
 With these initial and boundary conditions, we perform time 
integration of equations (18)--(23) until a quasi-steady solution is obtained.

\subsection{Time variations of the luminosity and the mass-outflow rate}
We show here the results of model A and B with $\beta=10^{-3}$.
The total luminosity $L_{\rm T}$ is a good measure to check whether a steady state 
of the accretion flow is attained and is given by
 $ L_{\rm T} = \int (q_{\rm br} + q_{\rm syn}) {\rm d} V$.  
The mass-outflow rate $\dot M_{\rm out}$ is here given as the total flow rate of the outflow
 at the outer boundary. 
In model A, a shock is at first formed at $r \sim 15R_{\rm g}$ on the equatorial
plane, expands outwards gradually and finally moves up and down irregularly around $r \sim 50 R_{\rm g}$. 
Due to the modulation of the shock position, the total luminosity varies by a factor of 2
and the averaged one is $4.0\times 10^{36}$ erg s$^{-1}$.
The mass-outflow rate also varies by a factor of $\sim$ 4 and its averaged mass-ouflow rate is as high as
 $\sim 1.0 \times 10^{20}$ g s$^{-1}$.  
 About half of the input accreting matter is ejected as the wind. 
 In model B, the shock wave appears in the inner region of $r \sim 19R_{\rm g}$  which is a predicted 
 value from the analysis in the 1D adiabatic case. 
 Then, the shock moves outwards gradually, recedes inwards after attaining to a position of $\sim 140 R_{\rm g}$
 and finally continues to oscillate irregularly between the radii of 45 and 65 $R_{\rm g}$.
 During the shock evolution, another weak shock appears inside the outer shock, disappears interacting with
 the outer shock and the inner edge and the processes are repeated irregularly. 
 The shock phenomena cause the complicated modulations of the luminosity and the mass-outflow rate.
 Fig.~3 shows the time variations of the total luminosity $L_{\rm T}$, the luminosity of the synchrotron radiation
 $L_{\rm syn}$, the mass-outflow rate $\dot M_{\rm out}$, 
 the mass-inflow rate $\dot M_{\rm in}$ at the inner edge and the shock position $R_{\rm s}$ 
 on the equatorial plane in model B. The total luminosity varies by a factor of 2 with time-scales of
 hours to days and the averaged luminosity is $2.5\times 10^{35}$ erg s$^{-1}$. 
 The luminosity of the synchrotron radiation
 is $1.5\times 10^{33}$ erg s$^{-1}$ and varies largely by  a factor of $\sim 10$.
 The mass-outflow rate varies intermittently by a factor of 1.5.
 and the averaged value at later phases is $1.0\times 10^{19}$ g s$^{-1}$ which is  several percent of the
 input matter. Accordingly, in this model, most of the input accreting matter falls into the event 
 horizon through the inner edge.
 
The origin of the oscillating shock appearing in the present models is entirely attributed to the range of 
 parameters of the specific energy and the specific angular momentum used here. 
 The oscillatory shock is necessarily triggered if the specific angular momentum $\lambda$
 and the specific energy $\epsilon$ belong to or are located just nearby in the range of parameters
 responsible for a stationary shock in the rotating inviscid and adiabatic accretion flow, as is found in the previous paper.
 
 Sgr A* possesses a quiescent state and a flare state. 
The flares and their quasi-periodicity of Sgr A* have been detected in multiple wavebands from radio, sub-millimetre, IR to X-ray \citep{b8,b9,b5,b10,b11,b23,b28,b29}.
 The amplitudes of the flare at radio, IR and X-ray are by factors of below 1/2, 1--5 and 45 or even higher, respectively, a few times per day. 
The rapid variability in the radio band on time-scales of a few seconds to hours has been observed \citep{b29}.  
These flare phenomena are interesting in relation to the time variations of
the luminosity. 
Actually, the modulation amplitude $\sim 2$ of the luminosity in models A and B can explain well the observed amplitudes of the variability at radio to IR except for the amplitude as high as $\sim$ 45 at X-ray.

 \begin{figure}
     \includegraphics[width=86mm,height=66mm]{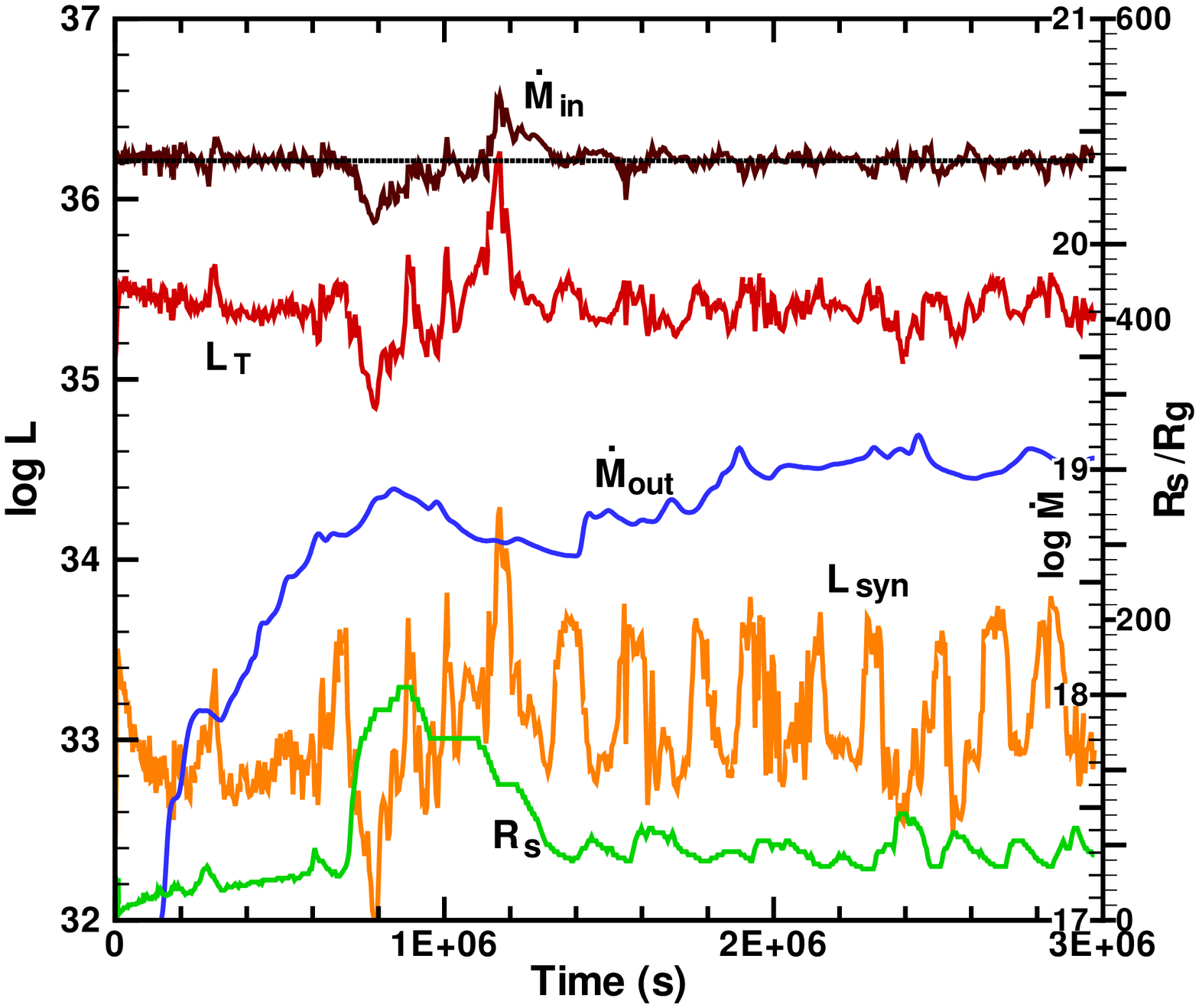}
     \caption{Time evolutions of total luminosity $L_{\rm T}$ (erg s$^{-1}$), the luminosity of the 
     synchrotron radiation $L_{\rm syn}$ (erg s$^{-1}$), 
       mass-outflow rate $\dot M_{\rm out}$ (g s$^{-1}$), mass-inflow rate $\dot M_{\rm in}$ (g s$^{-1}$)
      at the inner edge and 
     shock position $R_{\rm s}$ (dotted line) on the equatorial plane in model B with $\beta=10^{-3}$,
     where the dotted line shows the input accretion rate 2.44 $\times 10^{20}$ (g s$^{-1}$). }  
     \label{fig3}
 \end{figure}

\subsection{Structure of the flow and the shock}

Fig.~4 shows the profiles of density $\rho$ (g ${\rm cm}^{-3}$),
ion temperature $T_{\rm i}$ (${\rm K}$), electron temperature $T_{\rm e}$, radial velocity $v$, 
 Mach number $M_{\rm a}$ on the equatorial plane
  versus radius at $t=2.8\times 10^6$ s for model B with $\beta=10^{-3}$. 
 The outer shock and the weak inner shock are found at $r= 48$ and $10 R_{\rm g}$, respectively.
 The upstream electron temperatures in front of the outer shock equal to the ion temperatures but the 
  downstream electron temperatures become to be lower than the ion temperatures in the post-shock region. 
  The upstream electron adiabatic index $\gamma_{\rm e}$ (=1.6) changes into 4/3 behind the shock.
  It is because the relation of  the post-shock electron temperature $ > T_{\rm c} > $ the pre-shock
  electron temperature happens to be established here that the electron temperatures begin to deviate 
  from the ion temperatures at the shock front. 
  Both the electron and ion temperatures in the post-shock region are higher than those of the 1D
  stationary solution in Fig.~1 and the total luminosity $ 2.5\times 10^{35}$ erg s$^{-1}$ at this phase
  is high by a factor of 3 compared with $ 8.6\times 10^{34}$ erg s$^{-1}$ in the 1D solution.

  \begin{figure}
     \includegraphics[width=86mm,height=76mm]{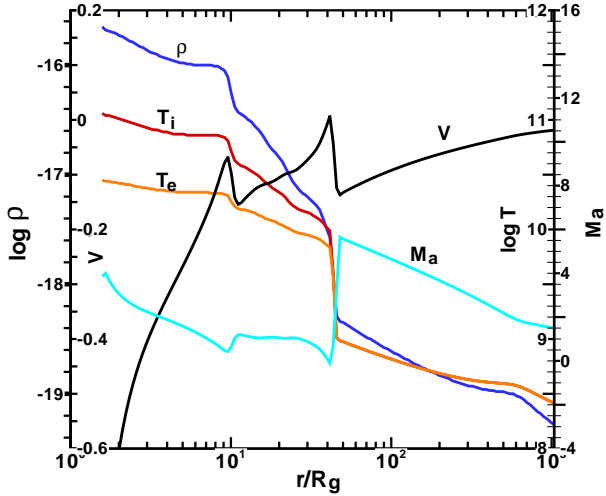}
     \caption{Profiles of density $\rho$ (g cm$^{-3}$), ion temperature 
     $T_{\rm i}$ (K), electron temperature $T_{\rm e}$ (K), radial velocity $v$ and
     Mach number $M_{\rm a}$ of the radial velocity
      on the equatorial plane versus radius at $t = 2.8\times 10^6$ s 
     for model B with $\beta=10^{-3}$. The outer shock and the inner weak shock are found at $r$ = 48
     and 10 $R_{\rm g}$, respectively.}
     \label{fig4}
  \end{figure}
  \begin{figure}
  
     \includegraphics[width=86mm,height=76mm]{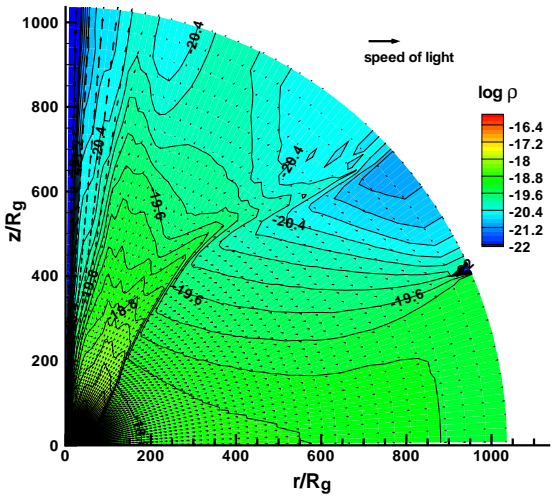}
     \caption{Contours of density $\rho$ $({\rm g}$ ${\rm cm}^{-3})$ with 
     velocity vectors at $t = 2.8\times 10^6$ s of model B. 
     The shock wave extends upwards obliquely at $r/R_{\rm g}\sim$ 48 
     on the equatorial plane.
     }  
     \label{fig5}
 
   \end{figure}

  Fig.~5 shows the contours of density $\rho$ 
 (g cm$^{-3}$) with velocity vectors at the same phase as Fig.~4 of model B.
The outer shock extends obliquely upwards at $r \sim 48 R_{\rm g}$ 
on the equatorial plane.
 The initial accreting matter at the outer boundary supersonically
falls down towards the gravitating centre and is decelerated at the shock 
front, enhancing the density and the temperature, and becomes subsonic.
The post-shock region with high densities and high temperatures 
results in a strong outward pressure-gradient force  along the 
$z$-axis and a part of the accreting matter deviates from the disc flow 
and escapes as the wind flow. The mass-outflow begins only after the matter
undergoes the shock compression.
The wind velocities become $0.01$--$0.05c$ at the outer boundary except for a 
narrow funnel region along the $z$-axis where the flow is relativistic as $\sim$
 0.2 -- 0.4$c$.

 \begin{figure}
     \includegraphics[width=86mm,height=66mm]{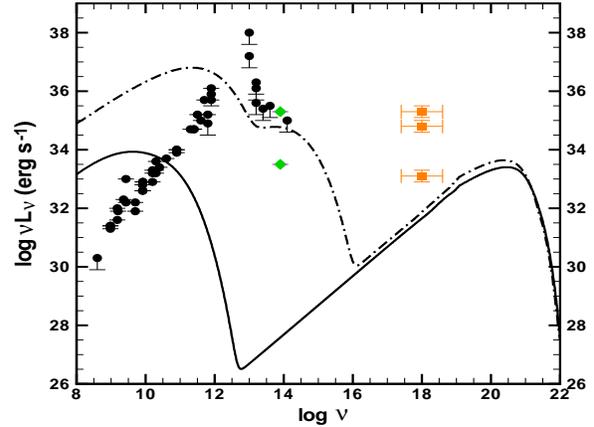}
     \caption{Energy spectra $\nu L_{\rm \nu}$ (${\rm erg}$
      ${\rm s}^{-1}$) of model B with $\beta=10^{-3}$ (solid line) at $t=2.8\times 10^6$ s and initial 
 model B with $\beta=1.0$ (dash--dotted line), where the symbols of filled circle, diamond and  cross
 are the observed data \citep{b15,b1,b8,b20,b9}}.   
     \label{fig6}
 \end{figure}

 Fig.~6 shows the energy spectra calculated from model B with $\beta=10^{-3}$ (solid line) at $t=2.2\times 10^6$ s,
 where the symbols of filled circle, diamond and cross show the observed points \citep{b15,b1,b8,b20,b9}.
 The submillimetre and X-ray bumps in these spectra are originated from 
 the synchrotron cooling by thermal electron and the bremsstrahlung cooling,
 respectively, whose formulas are prescribed in \citet{b22} and \citet{b18}.
 The peak intensities in the sub-millimetre bumps in this model are
 low and the peak frequencies lie in longer wavelength compared with the 
 observed spectra at the quiescent state of Sgr A*.
 To the contrary, the intensities in the X-ray band ($\sim 1$ keV) are lower than the observed one 
 but the X-ray spectra seem to be rather hard.
 In Table 2, we showed that the synchrotron emission increases with increasing $\beta$ and the total luminosity
 is $\sim 10^{36}$ erg s$^{-1}$  in model B with $\beta \sim 1$. 
 For the comparison with the above spectra, we show the energy spectra of model B with $\beta=1.0$  
 (dash-dotted line) in Fig.~6.  The spectra for $\beta=1.0$ in the radio to IR bands 
 seem to be improved somewhat but are not sufficient to explain the observed one.
 The present simple model cannot reproduce the detailed spectra of Sgr A* as are 
 successfully found in the ADAF models \citep{b26,b27} but it is useful to examine the basic 
 scenario of the low angular momentum flow model.

\section{Summary and discussion}
 Following the previous paper based on the analysis of the angular momentum of 
 the accretion flow around Sgr A* by \citet{b13},
we examined a low angular momentum flow model for Sgr A*
using 1D stationary and  2D time-dependent hydrodynamical calculations under the two-temperature model.
Adopting the model parameters of the specific angular momenta $\lambda$ of 1.35 and 1.68 
 and the input accretion rates of $\dot M_{\rm input} = 4.0\times 10^{-6}{\rm M_{\odot}}$ yr$^{-1}$, 
 we summarize the results as follows.

 (1) As long as the ratio $\beta$ of the magnetic energy
   density to the thermal energy density is taken to be as $\beta \le 1$,
 the radiative energy loss and gain due to the Coulomb collisions,
 the bremsstrahlung and the synchrotron emissions are negligible on the dynamics of the flow and 
 the two-temperature model is ascribed to the adiabatic two-temperature model.

 (2) The different temperatures of ions and electrons are caused by the different adiabatic indices of ions
 and electrons which depend on their temperature states under the relativistic regime.
  The total luminosity increases with increasing $\beta$ and results in $\sim 10^{35}$ -- $10^{36}$ 
 erg s$^{-1}$ for $\beta=10^{-3}$ -- 1.0 which is comparable to the observed luminosity of Sgr A*.

 (3) Due to the constant low angular momentum flow, the irregularly oscillating shocks are formed
 in the inner region, as is found in the previous paper, and the luminosity and the mass-outflow rate are
  modulated by a factor of 2 -- 3 and 1.5 -- 4, respectively, on time-scales of hours to days.

 (4) The time variability may be related to the flare phenomena of Sgr A*.

 Further evidence of the low angular momentum of the accretion
 flow around Sgr A* would confirm the validity of the model.
 If the specific angular momentum of the accretion flow around Sgr A* is actually as small as $\lambda$=1.35 --1.68,
 the time variabilities of the total luminosity found in models A and B  may explain 
 the quiescent and flare states of Sgr A* at the radio to IR bands except for X-rays. 
 However, there remain important problems of the too high X-ray variability and the time lag between the
 radio and X-ray flares in Sgr A*.
 In stead of this scenario for the flare activity of Sgr A*, we may interpret alternative one for the quiescent
  and flare states of Sgr A*, somewhat on the analogy of the temporal sequence of accretion disc--jet
  coupling observed in the X-rays, IR and radio wavelengths of the microquasar GRS 1915+105 \citep {b12}.
  Then we regard the total emission obtained in the models as the permanent emission of the quiescent state
 but consider that the emission variabilities in Sgr A* are caused by the intermittent outflow found 
 in the present models, where the mass-outflow rates are as high as $2\times 10^{-7}$ -- $2\times 10^{-6}\;{\rm M_{\odot}}$yr$^{-1}$ and vary by a factor of 1.5 -- 4 and the outflow escapes as the high-velocity
  jets with 0.01 -- 0.2$c$ along the rotational axis. 
 We speculate here that the intermittent high-velocity jets propagate outwards, interact with the 
 surrounding matter and excite the matter as a hot emitter.
 The emission from the hot matter may be observed as an amplified flare in X-rays. After that,  the
 high-velocity jets expand furthermore, cool down adiabatically, interact again with the distant interstellar 
 cloud and emit radio emission which is observed as a radio flare with a time lag.
 In the present models, we have treated only the inviscid accretion flow. If the accretion flow is actually 
 viscous flow in Sgr A* system, the present result of the low angular momentum flow model will be changed 
 largely and the application of the shock phenomena to the time variability of Sgr A* will be invalid
 because of no shock formation.
 The present low angular momentum flow model is also simple and cannot reproduce the detailed observed spectra
 of Sgr A*.
 For the spectral fitting with the observations, we need to treat the physics of the synchrotron
 and inverse Compton emission by the thermal and non-thermal electrons. 
 A further magneto-hydrodynamical examination of the flow with the low angular momentum including the above
 physics should be pursued in the future.

\section*{Acknowledgements}

I thank an anonymous referee for useful suggestion of the adiabatic index effects of ions and electrons
on the two--temperature model.

\label{lastpage}

\end{document}